\documentclass[12pt,preprint]{aastex}

\shorttitle{Search for Dark Matter with X-ray}
\shortauthors{Jeltema \& Profumo}

\begin{document}

\title{Searching for Dark Matter with X-ray\\ Observations of Local Dwarf Galaxies}

\author{T.~E.~Jeltema}
\affil{Morrison Fellow, UCO/Lick Observatories, 1156 High St., Santa Cruz, CA 95064, USA}
\email{tesla@ucolick.org}

\author{S.~Profumo}
\affil{Department of Physics and Santa Cruz Institute for Particle Physics\\ University of California, 1156 High St., Santa Cruz, CA 95064, USA}
\email{profumo@scipp.ucsc.edu}

\begin{abstract}
A generic feature of weakly interacting massive particle (WIMP) dark matter models is the emission of photons over a broad energy band resulting from the stable yields of dark matter pair annihilation. Inverse Compton scattering off cosmic microwave background photons of energetic electrons and positrons produced in dark matter annihilation is expected to produce significant diffuse X-ray emission. Dwarf galaxies are ideal targets for this type of dark matter search technique, being nearby, dark matter dominated systems free of any astrophysical diffuse X-ray background. In this paper, we present the first systematic study of X-ray observations of local dwarf galaxies aimed at the search for WIMP dark matter. We outline the optimal energy and angular ranges for current telescopes, and analyze the systematic uncertainties connected to electron/positron diffusion. We do not observe any significant X-ray excess, and translate this null result into limits on the mass and pair annihilation cross section for particle dark matter. Our results indicate that X-ray observations of dwarf galaxies currently constrain dark matter models at the same level or even more strongly than gamma-ray observations of the same systems, although at the expenses of introducing additional assumptions and related uncertainties in the modeling of diffusion and energy loss processes. The limits we find constrain portions of the supersymmetric parameter space, particularly if the effect of dark matter substructures is included. Finally, we comment on the role of future X-ray satellites (e.g. Constellation-X, XEUS) and on their complementarity with GLAST and other gamma-ray telescopes in the quest for particle dark matter.
\end{abstract}


\section{Introduction}

The fundamental nature of dark matter is at present unknown. It is widely believed that dark matter is in the form of a particle lying outside the ranks of the Standard Model of particle physics. An attractive possibility is that the New Physics sector hosting the dark matter particle is connected to the electro-weak scale, soon to be explored with the Large Hadron Collider. Motivations in support of this possibility include the fact that several models for new, electro-weak scale physics encompass particles that have all the microscopic features necessary to be the dark matter (this is the case for supersymmetry \citep[see e.g.][]{marcsreview}, models with universal extra-dimensions \citep[see e.g.][]{ued}, little Higgs models \citep[see e.g.][]{lhm}, and many others \citep[see e.g.][]{hooperreview}); in addition, weakly interacting particles with electro-weak scale mass (or WIMPs, for weakly interacting massive particles) possess the virtue that an order of magnitude estimate of their thermal relic abundance falls in the same range as the density of cold dark matter inferred from various cosmological observations \citep{wmap}. WIMPs might have once been in thermal equilibrium in the early universe, eventually freezing-out when the rate of pair-annihilation fell below the Hubble expansion rate: particle dark matter, in this scenario, would be one more ``{\em thermal relic}'', similar to the cosmic microwave background photons or the light elements produced during Big Bang Nucleosynthesis.

The fact that, occasionally, WIMPs can still pair-annihilate in the present universe fostered a wide array of indirect particle dark matter searches: the pair annihilation of dark matter is expected to yield, for instance, a significant amount of energetic antimatter (including GeV positrons, antiprotons and antideuterons \citep[see e.g.][]{1999PhRvD..59b3511B,2004PhRvD..69f3501D,amprofumo,dbar,2007PhRvD..76h3506B,2007PhRvD..75h3006B}, and gamma rays, from various radiative processes and from the decay of neutral pions resulting from the hadronization of particles in the pair-annihilation final state. Of interest to us here is the fact that particle dark matter annihilation also produces a population of energetic electrons and positrons from, for instance, charged pion, muon, gauge and Higgs boson decays, and, possibly, prompt production. The injection spectrum depends on the details of the particle dark matter model, but since electrons and positrons ($e^+e^-$) are produced at a center-of-mass energy corresponding to twice the particle dark matter mass, assumed here to be around the electro-weak scale, many of them will have energies at or above a GeV. These energetic $e^+e^-$ populate dark matter halos in any generic WIMP model, with densities which depend on both the dark matter density profile and the WIMP pair annihilation rate. Electrons (and positrons) diffuse, loose energy and produce secondary radiation through various mechanisms. In the presence of magnetic fields they emit at radio wavelengths via synchrotron radiation. Inverse Compton (IC) scattering off target cosmic microwave background photons and background light at other frequencies gives rise to a broad spectrum of photons, stretching from the extreme ultra-violet up to the gamma-ray band. A further, typically subdominant contribution to secondary photon emission results from non-thermal bremsstrahlung, i.e. the emission of gamma-ray photons in the deflection of the charged particles by the electrostatic potential of intervening gas.

The multi-wavelength emission from dark matter annihilation was studied in detail in the seminal works of \cite{baltzwai} for galactic dark matter clumps, in \cite{colacoma} for the case of the Coma cluster and in \cite{coladraco} for the dwarf spheroidal galaxy Draco. Other recent studies include an interpretation of the significant non-thermal X-ray activity observed in the Ophiuchus cluster in terms of IC scattering of dark-matter produced $e^+e^-$ \citep{ophiuchus}, an analysis of the broad-band dark-matter annihilation spectrum expected from the Bullet cluster \citep{colabullet} and from the galactic center region \citep{ullioregis}. In addition, radio emission from $e^+e^-$ produced in dark matter annihilation was considered as a possible source for the ``WMAP haze'' in the seminal paper of \cite{haze1}, and subsequently analyzed in detail in \citet{haze2}, \citet{haze3} and \citet{2008arXiv0801.4378H}. Other studies have also previously addressed synchrotron radiation induced by dark matter annihilation \citep[e.g.][]{gondolo,bertone0101134,aloisio0402588}.

Among the possible targets for the observation of an astronomical signature of dark matter annihilation, local dwarf spheroidal (dSph) galaxies stand out as excellent candidates for several reasons. First, unlike the galactic center region or galaxy clusters, no significant diffuse radio, X-ray or gamma-ray emission is expected: the gravitational potential well of dSph galaxies is too shallow for them to host any sizable thermal bremsstrahlung emission at X-ray frequencies, and, more importantly, the gas densities appear to be extremely low \citep[see e.g.][]{1998ARA&A..36..435M}. Second, dSph are the most dark matter dominated known systems, and, with the exception of our own Milky Way, they are the closest known bound dark matter systems. Unlike a signal from the galactic center region or from a nearby cluster, a diffuse X-ray or radio emission from a nearby dSph galaxy would likely not have an obvious astrophysical counterpart that could fake a dark-matter induced emission. The cross correlation of diffuse emission from one of the nearby dSph galaxies with, for instance, point-like emission at gamma-ray frequencies detected with GLAST (unlike secondary emission from $e^+e^-$, species which undergo spatial diffusion, gamma rays trace the dark matter density profile squared), and a study of the spectral features at various wavelength could potentially lead not only to the conclusive detection of dark matter, but even to the identification of the particle mass and some of its microscopic particle properties. The spectral shape of the broad-band emission depends in fact on the dominant Standard Model final state into which the dark matter particle annihilates, and this is in turn determined by the WIMP particle model.

In this paper, we present the first systematic study of archival X-ray data on local dSph galaxies with the aim of detecting a signal from dark matter annihilation. After introducing the physics of $e^+e^-$ production from dark matter annihilation and of subsequent diffusion and energy loss, we present in Sec.~\ref{sec:multiw} a few examples of multi-wavelength spectra, emphasizing and analyzing the role of systematic uncertainties connected with the dark matter particle model and with the diffusion setup. We then present our data reduction and analysis in Sec.~\ref{sec:data}, motivating our choice for the energy range and angular region. We present limits on the X-ray diffuse emission from the dSph galaxies Fornax, Carina and Ursa Minor in Sec.~\ref{sec:xraylimt}. Sec.~\ref{sec:constraints} is devoted to a study of how our null result limits particle dark matter models. Our summary and a discussion of the role of future X-ray detectors in particle dark matter searches, including their connection with gamma-ray telescopes, are presented in the final Sec.~\ref{sec:discussion}.

\section{The Multi-wavelength Spectrum from Dark Matter Annihilation}\label{sec:multiw}

The starting point for the computation of the multi-wavelength spectrum resulting from dark matter annihilation is the identification of the relevant source function. Given a stable species $i$ and a position $\vec{r}$, the source function $Q_i(\vec{r},E)$ \citep[we are using here the same notation of][]{colacoma} gives the differential number of particles $i$ per unit time, energy and volume element produced at $\vec{r}$ and with an energy $E$:
\begin{equation}
Q_i(\vec{r},E)=\langle\sigma v\rangle_0\sum_f\frac{{\rm d}N_i^f}{{\rm d}E}(E)\ B_f\ {\cal N}_{\rm pairs}(\vec{r}),
\end{equation}
where $\langle\sigma v\rangle_0$ is the WIMP pair annihilation rate at zero temperature, and the sum is over all kinematically allowed Standard Model annihilation final states $f$ (for instance, quark-antiquark, $W^+W^-$, lepton-antilepton etc.), each weighed with a branching ratio $B_f$ and producing a spectral distribution ${\rm d}N_i^f/{\rm d}E$, after prompt production or decay and fragmentation into the stable particle $i$. Finally, ${\cal N}_{\rm pairs}(\vec{r})$ is the number of dark matter particle pairs per volume element squared, which for the case of a smooth dark matter distribution $\rho_{\rm DM}(\vec{r})$ is given by
\begin{equation}
{\cal N}_{\rm pairs}(\vec{r})=\frac{\rho^2_{\rm DM}(\vec{r})}{2\ m_{\rm DM}},
\end{equation}
where we indicate with $m_{\rm DM}$ the mass of the dark matter particle.

In the case of gamma rays, since photons propagate on straight lines, the flux from prompt emission in a given direction is simply given by the integral of the appropriate source function along the line of sight,
\begin{equation}\label{eq:gr}
\frac{{\rm d}N_\gamma}{{\rm d}E_\gamma}=\int_{\rm l.o.s.}\ {\rm d}l\ Q_\gamma(E_\gamma,\vec{r}(l)).
\end{equation}
The quantity in Eq.~(\ref{eq:gr}) is then integrated over the angular region over which the signal is observed, and convolved with the angular dependent instrumental sensitivity of the gamma-ray telescope under consideration.

The treatment for electrons and positrons is complicated by diffusion and energy loss processes. We model these with a diffusion equation of the form
\begin{equation}\label{eq:trans}
\frac{\partial}{\partial t}\frac{{\rm d}n_e}{{\rm d}E}=\nabla\Big[D(E,\vec{r})\nabla\frac{{\rm d}n_e}{{\rm d}E}\Big]+\frac{\partial}{\partial E}\Big[b(E,\vec{r})\frac{{\rm d}n_e}{{\rm d}E}\Big] + Q_e(E,\vec{r}),
\end{equation}
where ${\rm d}n_e/{\rm d}E$ is the electron and positron spectrum, $D(E,\vec{r})$ is the diffusion coefficient and $b(E,\vec{r})$ is the energy loss term. Analytical solutions to the equation above exist in the equilibrium regime \citep[see e.g. App.~A in][]{colacoma}, as long as the spatial dependence of both the diffusion coefficient and the energy loss term are dropped, and spherical symmetry in the dark matter distribution is assumed (in the present study we make the same hypotheses). The dependence of the diffusion coefficient on energy is assumed to be a power law of the form
\begin{equation}
D(E)=D_0\ \left(\frac{E}{\rm 1\ GeV}\right)^\gamma.
\end{equation}
Very little is known about diffusion in systems of the type of interest here, i.e. dSph galaxies. This forces us to make educated guesses about $D_0$ and $\gamma$. The best known system as far as cosmic ray propagation is concerned is by all means our own Galaxy. \cite{2004PhRvD..69f3501D} analyzed data on cosmic ray
fluxes in the Milky Way, in particular ratios of primary to secondary
species. The outcome of their analysis was to determine the preferred
values for the parameters $D_0$ and $\gamma$, in the framework of a diffusion
setup similar to the one outlined here.  Their median values are $D_0=1.1\times 10^{27}\ {\rm cm}^2/{\rm s}$ and $\gamma=0.7$, which we employ here as reference values. Using the phenomenological ranges found by \cite{2004PhRvD..69f3501D}, we will also consider values of the parameter $0\le\gamma\le1$. 

The parameter $D_0$ is related to the size and scale of galactic magnetic field inhomogeneities; data on galaxy clusters indicate that larger systems feature larger values for $D_0$ than galactic-size systems. This can be understood by estimating the diffusion coefficient $D_0$ in the context of passive advective transport in a turbolent flow as the product of the turbolent velocity $V_L$ and of the turbolent injection scale $L$ \citep[see e.g.][]{2001ApJ...562L.129N,2004Ap&SS.289..307L}. If the turbolent velocity in clusters of galaxies is of the order of the velocity dispersion of galaxies (say $V_L\sim 1000$ km/s), and the injection scale is of the order of the galactic scale (say, $L\sim 20$ kpc), then the predicted $D_0\sim L V_L\sim 6\times 10^{30}\ {\rm cm}^2/{\rm s}$ is indeed close in value to what is observed in clusters \citep[][quote for instance $D_0\sim 6.22\times 10^{30}\ {\rm cm}^2/{\rm s}$ for the Hydra A galaxy cluster]{2003ApJ...582..162Z}. Scaling this estimate from cluster down to galactic scales, and using the reference value for $D_0$ from galactic cosmic ray data obtained by \cite{2004PhRvD..69f3501D} indicates, assuming $V_L$ for the galaxy is smaller by a factor 10 than in clusters, that the value of $L$ for galaxies is at least a factor 100 smaller than for clusters. Conservatively assuming that $L$ does not change switching from a Milky Way size galaxy to a dSph, the simple scaling in $V_L$ of the diffusion coefficient points to $D_0\sim10^{26}\ {\rm cm}^2/{\rm s}$, given that the velocity dispersion of the Milky Way is more than one order of magnitude larger than those observed in local dSph galaxies.

In this respect, we might expect values for $D_0$ smaller than the conservative Milky Way median value we use as a reference here. We will therefore also consider as an alternative value $D_0=10^{26}\ {\rm cm}^2/{\rm s}$, and we will discuss the dependence of the X-ray emission from dark matter pair-annihilation on both $D_0$ and $\gamma$ in Sec.\ref{sec:constraints}. 

Another important parameter for the diffusion model is the diffusion volume, where Eq.(\ref{eq:trans}) is solved, at the boundary of which free-escape boundary conditions are imposed. Following \cite{coladraco}, we assume a spherical diffusion zone, and a diffusion radius $r_h$ corresponding to twice the radius of the stellar component, typically a few kpc for local dSph galaxies \citep{1998ARA&A..36..435M}. This choice is again motivated by analogy with the picture for the Milky Way. The value of $r_h$ is also a crucial parameter for the computation of the X-ray emission from dark matter annihilation: the smaller the radius of the diffusion region, the larger the fraction of $e^+e^-$ escaping and diffusing away, hence the smaller the X-ray emission.

The energy loss term, which is assumed to be position-independent, includes the following terms \citep{colacoma}:
\begin{equation}\label{eq:enloss}
b(E)=b_{\rm IC}(E)+b_{\rm syn}(E)+b_{\rm coul}(E)+b_{\rm brem}(E),
\end{equation}
where
\begin{eqnarray}
\nonumber b_{\rm IC}(E)=b^0_{\rm IC}\left(\frac{E}{1\ {\rm GeV}}\right)^2 & \quad & b^0_{\rm IC}\simeq0.25\times 10^{-16}\ {\rm GeV}{\rm s}^{-1}\\
\nonumber b_{\rm syn}(E)=b^0_{\rm syn}\left(\frac{B}{1\ \mu{\rm G}}\right)^2\left(\frac{E}{1\ {\rm GeV}}\right)^2 & \quad & b^0_{\rm syn}\simeq0.0254\times 10^{-16}\ {\rm GeV}{\rm s}^{-1}\\
\nonumber b_{\rm coul}(E)=b^0_{\rm coul}\ n\ \left(1+\log(\gamma_e/n)/75\right) & \quad & b^0_{\rm coul}\simeq6.13\times 10^{-16}\ {\rm GeV}{\rm s}^{-1}\\
\nonumber b_{\rm brem}(E)=b^0_{\rm brem}\left(\log(\gamma_e)+0.36\right)\left(\frac{E}{1\ {\rm GeV}}\right) & \quad & b^0_{\rm brem}\simeq1.51\times 10^{-16}\ {\rm GeV}{\rm s}^{-1}
\end{eqnarray}
where $n$ indicates the thermal electron density and $\gamma_e=E/(m_e c^2)$. We set in what follows the average magnetic field to $B=1\ \mu$G, in concordance with radio observations at 5 GHz of dSph galaxies reported in \cite{klein}, and the thermal electron density to $n=10^{-6}\ {\rm cm}^{-3}$ \citep[see][]{colacoma}.

No measurements are available on the average magnetic field for the dSph galaxies under investigation here. Eq.~(\ref{eq:enloss}) indicates that with our reference choice for $B$, energy losses for energetic electrons and positrons ($E\gtrsim 1$ GeV) are dominated by Inverse Compton processes. Smaller values for the magnetic field would thus in no way affect our results. However, larger values for $B$ could be in principle allowed by available data \citep[see e.g.][]{klein}. Synchrotron losses dominate over IC for $B\gtrsim 4\ \mu$G. Above that value, we expect a suppression on the secondary X-ray emission from WIMP annihilation, since the density of energetic electrons and positrons are depleted by more efficient energy losses. Larger average magnetic fields would however also produce an intense radio emission, also from secondary processes related to WIMP annihilation. The results of \cite{coladraco} indicate that, in the case of the Draco dSph, for $B\gg 1\ \mu$G, the synchrotron emission originating from WIMP annihilation would violate VLA limits on radio emission \citep[][]{klein}. The thermal electron density only affects the strength of Coulomb losses (see Eq.~(\ref{eq:enloss})), and thus the low energy part ($E\ll 1$ GeV) of the electron-positron equilibrium spectrum. Larger values for $n$ would thus only affect the low energy end of the IC emission in the multi-wavelength spectrum, well below the X-ray regime of interest here. In addition, $n\gg 10^{-6}\ {\rm cm}^{-3}$ would lead to some significant thermal bremsstrahlung emission, potentially in conflict with the limits on X-ray emission we present below.

After specifying the dark matter density profile and the $e^+e^-$ injection spectrum, Eq.(\ref{eq:trans}) can be integrated to find the equilibrium distribution ${\rm d}n_e/{\rm d}E$. In turn, knowledge of the spatial and energy distribution of $e^+e^-$ allows us to compute the multi-wavelength secondary emission. For our purposes, and in the energy range of interest here, the dominant contribution comes from the up-scattering of CMB photons. While a contribution from starlight and background light at other frequencies is also expected, it is generically subdominant in the X-ray band.

The inverse Compton power is obtained by folding the number density of target photons $n(\varepsilon)$ (here, the CMB black body spectrum) with the IC scattering cross section:
\begin{equation}
P(E_\gamma,E)=cE_\gamma\int{\rm d}\varepsilon\ n(\varepsilon)\sigma(E_\gamma,\varepsilon,E),
\end{equation}
where $E=\gamma_e m_e c^2$ is the $e^+e^-$ energy, $\varepsilon$ is the target photon energy, $E_\gamma$  is the energy of the up-scattered photon, and $\sigma$ indicates the Klein-Nishina formula (the full relativistic QED form of the Thomson scattering cross-section). Folding the IC power with the $e^+e^-$ equilibrium distribution, we get the local emissivity
\begin{equation}
j(E_\gamma,\vec{r})=\int\ {\rm d}E\ \left(\frac{{\rm d}n_{e^-}}{{\rm d}E}+\frac{{\rm d}n_{e^+}}{{\rm d}E}\right)\ P(E_\gamma,E).
\end{equation}
Finally, the integrated flux density spectrum, as in the case of gamma rays, is given by the angular and line of sight integral of the expression above (see Eq.~(\ref{eq:gr})).

The last ingredient needed to actually compute the broad band spectrum of dark matter annihilation is to specify the particle dark matter model. Up to a normalization factor depending on the dark matter density profile and on the dark matter pair annihilation rate, all that matters as far as the particle dark matter model is the mass $m_{\rm DM}$ and the set $\{B_f\}$ of branching ratios into given Standard Model final states. For simplicity and to allow comparison with other studies, we choose particle dark matter models with branching ratio 1 into a certain final state (i.e. always annihilating into the same Standard Model final state). We choose as a reference model a 100 GeV WIMP pair-annihilating into a $b \bar b$ pair (the resulting spectrum for other quark-antiquark final states is very similar to this one). Our choice is also motivated by the fact that many supersymmetric models feature this particular final state as the dominant one \citep[examples include the bulk and funnel regions of minimal supergravity, and generic bino-like models with large $\tan\beta$, see e.g.][]{hooperreview}. 

\begin{figure}
\mbox{\hspace*{-0.5cm}\includegraphics[width=8.5cm]{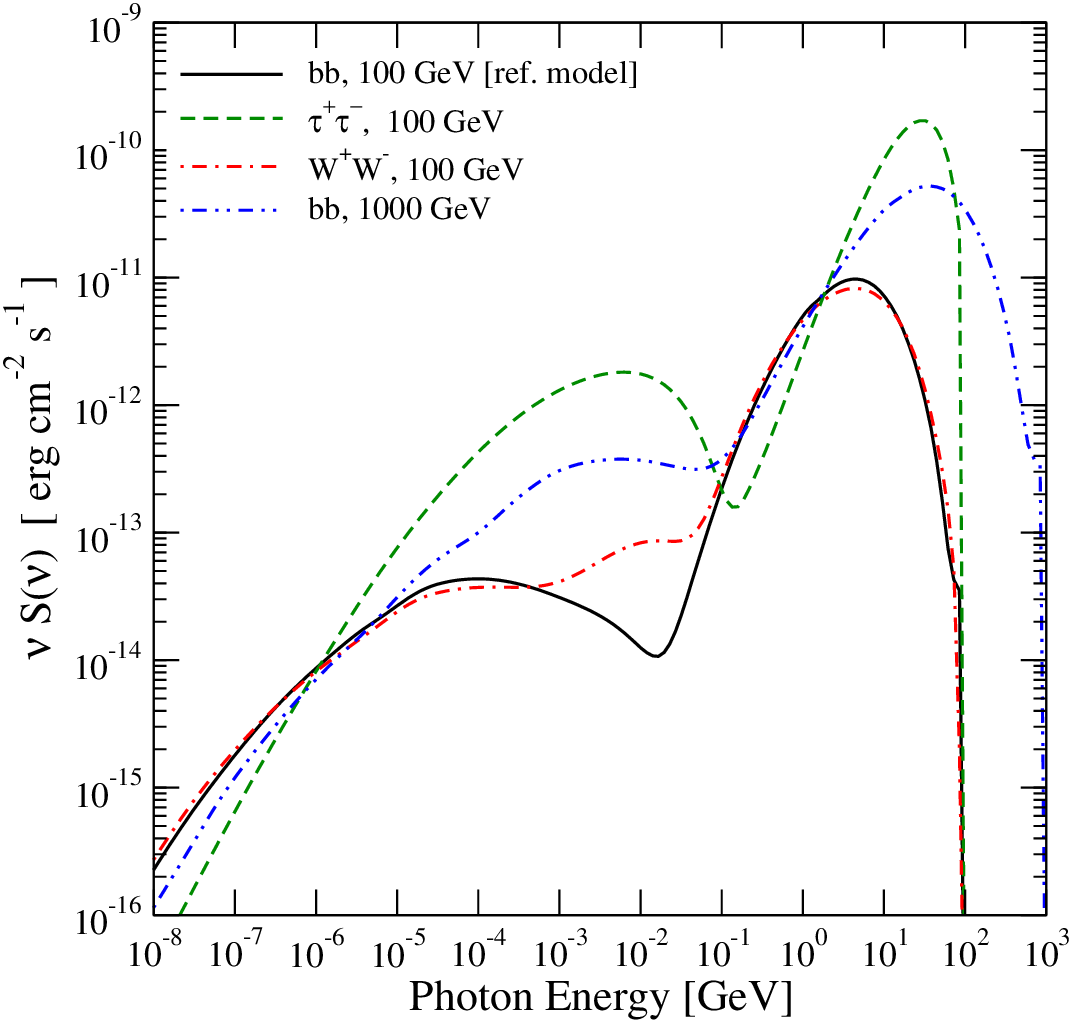}\qquad\includegraphics[width=8.5cm]{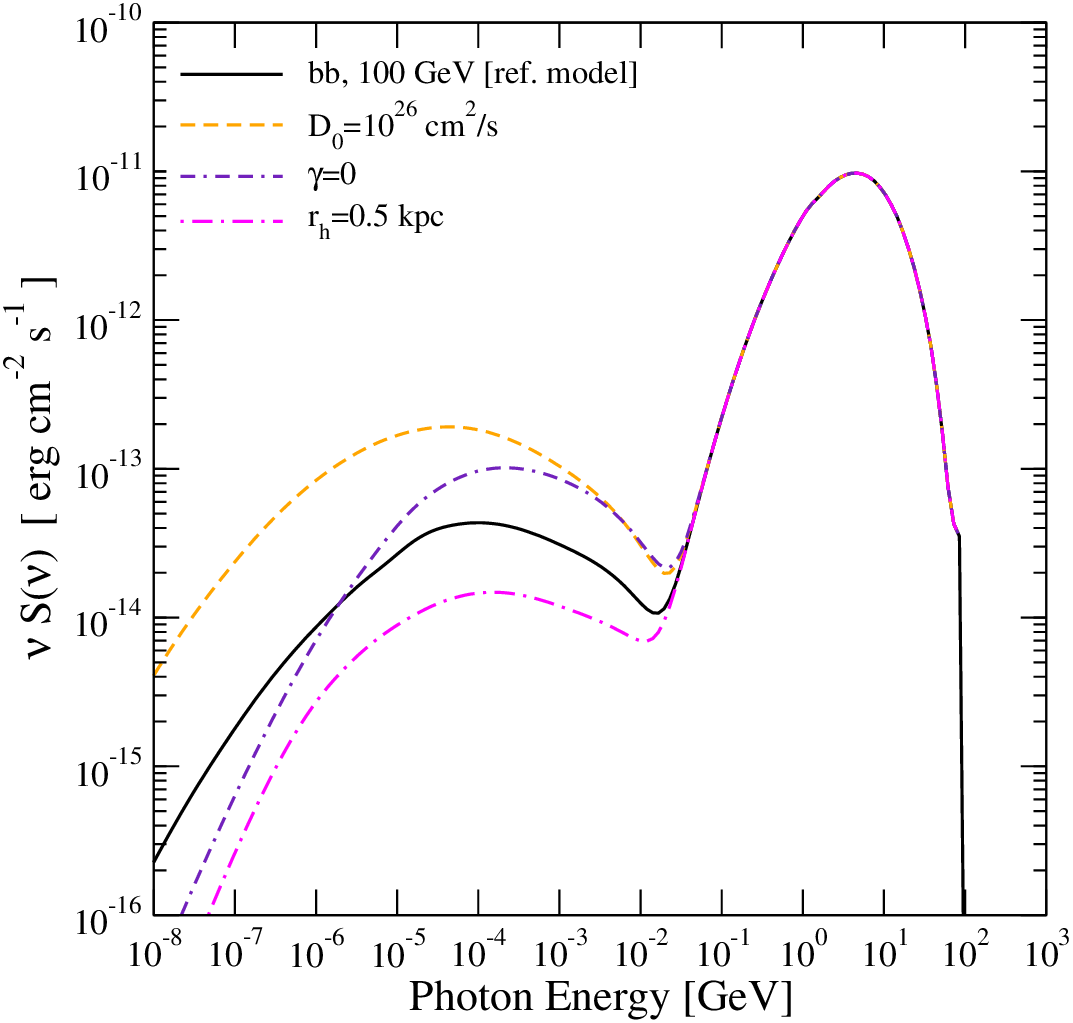}}
\caption{Left: The multi-wavelength spectrum (spectral energy distribution) of various particle dark matter models (the reference 100 GeV WIMP annihilating into $b\bar b$, two setups also with a mass of 100 GeV but pair-annihilation final states $\tau^+\tau^-$ and $W^+W^-$, and a model with the same final state as the reference model, but a mass of 1 TeV). Right: a comparison of the effect on the dark matter multi-wavelength spectrum of the reference model of changes to parameters in the diffusion model (a diffusion coefficient set to $10^{26}\ {\rm cm}^2/{\rm s}$ instead of the reference value $1.1\times 10^{27}\ {\rm cm}^2/{\rm s}$, a scaling with energy of the diffusion coefficient $\gamma=0$ instead of the reference value $\gamma=0.7$, and a radius for the diffusion region of $r_h=0.5$ kpc, versus the reference value of $2.4$ kpc).\label{fig:sed}}
\end{figure}

We show the spectral energy distribution (SED) for this particular dark matter particle model with a black line in both panels in Fig.\ref{fig:sed}, where we only show photon energies larger than 10 eV. We chose the normalization so that the integrated gamma-ray flux between 0.1 and 10 GeV is $10^{-8}$ photons per ${\rm cm}^2$ per s, over an angular region of one degree, which roughly corresponds to the EGRET point-source sensitivity \citep{egretpaper}. For reference, we use a Navarro-Frenk-White (NFW) dark matter density profile \citep{nfwpaper}
\begin{equation}\label{eq:nfw}
\rho_{\rm DM}(|\vec{r}|)=\rho_s\ \left(\frac{|\vec{r}|}{r_s}\right)^{-1}\ \left(\frac{|\vec{r}|}{r_s}+1\right)^{-2},
\end{equation}
where $|\vec{r}|$ indicates the distance from the center of the dSph galaxy, and where we set the scale radius $r_s=1$ kpc. Also, we set $r_h=2.4$ kpc.
In the left panel we assess how the particle physics model affects the dark matter annihilation SED. The dashed green line indicates the result for a model with the same 100 GeV mass, but annihilating into $\tau^+\tau^-$, a final state also motivated by supersymmetry (e.g. in the stau coannihilation region), which features a harder $e^+e^-$ injection spectrum, as well as a harder gamma-ray spectrum. A more abundant population of energetic electrons and positrons results in an IC spectrum which peaks at larger energies than for a softer injection spectrum, such as the reference model. In the soft X-ray band, however, while we find different spectral indices, the two models have a similar level of emission at a given gamma-ray luminosity. The dot-dashed red line indicates, again for $m_{\rm DM}=100$ GeV, the case of the $W^+W^-$ final state, found e.g. in Higgsino and Wino supersymmetric dark matter models. This case is intermediate between the two previous ones, and, compared to the reference model has a population of energetic electrons and positrons resulting from e.g. $W\to l\nu_l$, $l=e,\ \mu,\ \tau$, which is responsible for the bump in the IC emission in the MeV region. Changing the mass of the dark matter particle shifts the SED: a larger $m_{\rm DM}$ yields more energetic $e^+e^-$, as can be appreciated comparing the black and the double-dotted-dashed blue line ($m_{\rm DM}=1000$ GeV). Again, the X-ray region is however only mildly affected. Notice that a generic feature of the X-ray spectrum is a hard ($0.8<\alpha<1.5$) photon spectral index, harder than most astrophysical X-ray sources, particularly diffuse thermal emission.

In the right panel of Fig.~\ref{fig:sed} we study the impact of changing the parameters in the diffusion model on the dark matter annihilation SED. A smaller diffusion coefficient (orange dashed line, $D_0=10^{26}\ {\rm cm}^2/{\rm s}$) suppresses $e^+e^-$ diffusion and escape from the diffusive region, yielding a larger IC flux, particularly in the X-ray band of interest here. The choice of $D_0$ is therefore very significant in the computation of the expected X-ray flux; we will quantitatively analyze this statement in Sec.\ref{sec:constraints}, Fig.\ref{fig:compare}. A milder rigidity dependence in the diffusion coefficient (i.e. a smaller $\gamma$, set to zero for the double-dashed-dotted indigo line) leads, at fixed $D_0$, to a suppressed diffusion of the most energetic $e^+e^-$ (i.e.~the diffusion coefficient is smaller for $E>1$ GeV). As a consequence, we get an enhancement of the high-energy IC photons and a suppression at lower up-scattered photon energies. The result on the X-ray emission is a mild suppression. Finally, the magenta dot-dashed line indicates the effect of taking a diffusive region with a radius a factor 5 smaller than our benchmark choice. This corresponds to a diffusion volume more than a factor 100 smaller: in turn, this implies a much larger loss of $e^+e^-$ escaping the diffusion region, leading to a significant suppression of the electron/positron number density and, therefore, of the IC X-ray signal.

In summary, significant uncertainties affect the computation of the IC X-ray emission resulting from dark matter annihilation; while for a given diffusion model the differences in the soft X-ray band are rather mild, changing the parameters in the diffusion setup affects quite significantly both the normalization and (although less dramatically) the spectrum of the predicted SED; the reference diffusion setup chosen here gives rather conservative predictions for the X-ray flux; the outstanding features of the signal we are after are therefore (1) extended emission and (2) a hard spectral index.

\section{Data and Data Reduction}\label{sec:data}

Of the nearby Local Group dwarf spheroidal (dSph) galaxies, three have observations in the XMM-Newton archive: Ursa Minor observed for 52 ks between August and October 2005 (ObsIDs 0301690201, 0301690301, 0301690401, 0301690501), Fornax observed for 104 ks in August 2005 (ObsID 0302500101), and Carina observed for 42 ks in May 2004 (ObsID 0200500201).  Here we focus on XMM observations as its large effective area and large field-of-view are ideal for detecting faint diffuse emission, while its fairly small PSF allows us to exclude X-ray point source contamination from X-ray binaries and background AGN.  A couple of additional local dSphs have been observed with Chandra and Suzaku, but much of these data are not yet public.  All of the XMM observations were taken in Full Frame mode; for Ursa Minor and Carina the thin optical blocking filter was used, while for Fornax the medium filter was used.  Unfortunately, the background flare filtering, discussed below, reveals that three of the observations of Ursa Minor are highly contaminated with background flares.  In our analysis, we use only ObsID 0301690401 for this dwarf.  In addition, for the Fornax and Ursa Minor observations CCD6 on MOS1 was not available, but this CCD does not fall within the considered source region (see below).

The data were reduced using XMMSAS version 7.1.0, and all observations were reprocessed using the EPIC chain tasks.  For EMOS data, we use patterns 0-12 and apply the \#XMMEA\_EM flag filtering, and for EPN data, we use patterns 0-4 and flag equal to zero.  Due to the time variability in the spectra of background flares \citep{neva}, we filter for periods of high background in several energy bands.  We first apply a conservative cut on the high energy count rate matching the 2XMMp pipeline and the EPIC blank-sky event file filtering ($<2$ cts s$^{-1}$ for EMOS data and $<60$ cts s$^{-1}$ for EPN data).  This cut removes the most egregious flares. In the case of Carina, we found that the observations contained a number of milder flares that were not removed by this conservative cut, but never the less biased the background rate high, and for this dwarf we found it necessary to apply more stringent high energy count rate cuts of $<0.65$ cts s$^{-1}$ for EMOS data and $<3.5$ cts s$^{-1}$ for EPN data.  For all dwarfs, we then applied a 3$\sigma$ clipping to the source-free count rate in three energy bands, 0.5-2 keV, 2-5 keV, and 5-8 keV.  Here time bins (bin size of 100 secs) with rates more than 3$\sigma$ away from the mean are removed recursively until the mean is stable.  As noted above, the flare filtering excluded nearly all of the exposure time from three of the four Ursa Minor observations.  The final clean exposure times are listed in Table 1 along with the adopted dwarf central positions.

\section{Results}

We describe below the limits we obtain for the diffuse X-ray emission from the selected dSph galaxies (Sec.~\ref{sec:xraylimt}), and how these limits constrain particle dark matter models (Sec.~\ref{sec:constraints}).

\subsection{X-Ray Flux Limits}\label{sec:xraylimt}

\begin{figure}
\mbox{\includegraphics[width=8cm]{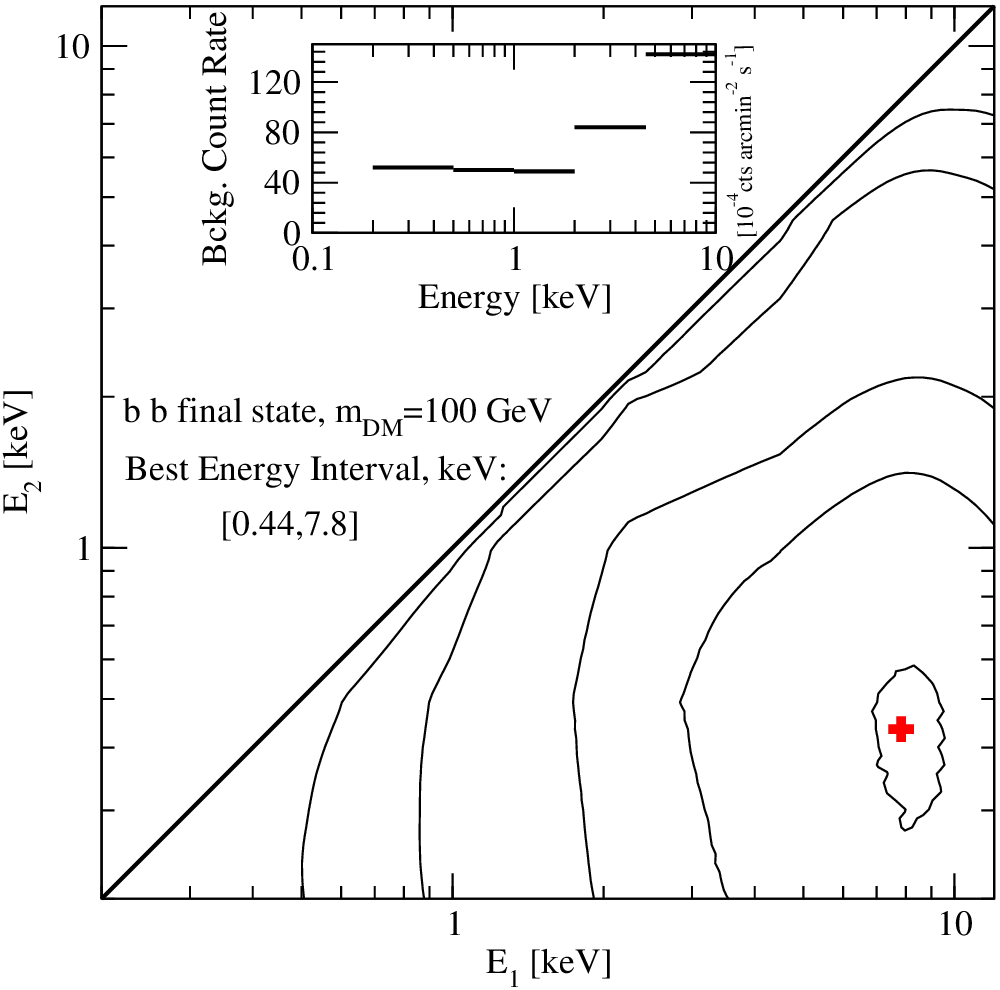}\qquad\includegraphics[width=8cm]{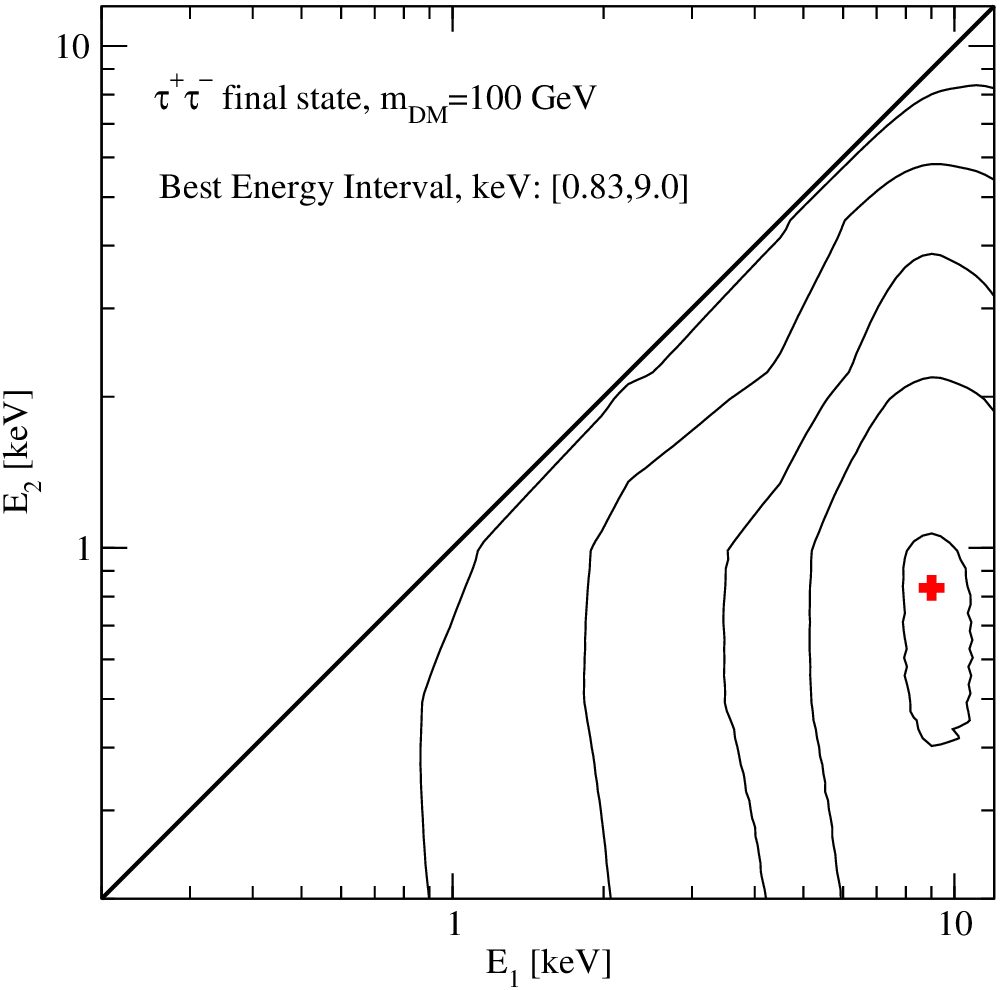}}
\caption{Curves at constant signal over the square root of the background in the plane defined by the upper and lower limits $E_1$ and $E_2$ of the energy interval over which the signal is integrated. The inset in the left panel shows the average XMM background count level we used. The left panel assumes the reference model (100 GeV, $b\bar b$ final state), while the right panel uses a model with a harder $e^+e^-$ injection spectrum. The red crosses indicate the $[E_1,E_2]$ interval giving the maximal S/N.\label{fig:optener}}
\end{figure}

Before comparing our model predictions to the actual X-ray data, we preformed a detailed study of the optimal energy and angular range where we expect the largest possible signal-to-noise ratio (S/N). Specifically, considering both the XMM-Newton affective area versus energy\footnote{http://xmm.esac.esa.int/external/xmm\_user\_support/documentation/uhb\_2.5/index.html} and the range of possible X-ray spectra resulting from the secondary emissions of $e^+e^-$ produced in WIMP annihilations, we find that the expected signal-to-noise peaks in roughly the 0.5-8 keV band. We illustrate this point in Fig.~\ref{fig:optener}, where we show contours of constant signal-to-noise in the plane defined by the lower energy limit $E_1$ and by the upper energy limit $E_2$. We integrate over the $[E_1,E_2]$ interval the signal obtained from WIMP annihilations (after diffusion, modeled according to the reference setup described above) and we divide by the square root of the background over the same energy interval. We show the XMM background count rate in the inset of the left panel \citep{2007A&A...464.1155C}. The S/N grows towards the red cross, which indicates the best values. In the left panel we assume a soft  $e^+e^-$ injection spectrum (for reference, a 100 GeV WIMP pair annihilating into $b\bar b$ pairs), while in the right panel we take a hard injection spectrum (100 GeV WIMP annihilating into $\tau^+\tau^-$ pairs). We verified that varying the WIMP mass doesn't affect the position of the best energy interval. Also, any combination of final states in the context of particle dark matter models motivated by beyond the Standard Model physics falls in between the two final states under consideration here. We obtain that with a soft injection spectrum the best energy range (giving the highest signal-over background) is [0.44,7.8] (all energies are in keV), while for a hard spectrum it is [0.83,9.0]. We decided to use the 0.5-8 keV range as the optimal X-ray band for XMM.

Similarly, we use the XMM sensitivity versus off-axis angle and the observed dark matter profiles of the dwarf galaxies \citep[see e.g.][]{2007PhRvD..75h3526S,1998ARA&A..36..435M} to explore the optimal source radius.  The total size of the dwarfs extends beyond the XMM field-of-view, but the expected X-ray flux decreases significantly with radius as the dark matter density drops as well as through the telescope vignetting. Neglecting the effect of diffusion, the maximal S/N is obtained for the smallest possible angular regions for profiles which diverge with $|\vec{r}|\to0$, such as the NFW profile. For cored profiles the best S/N (again in the limit of no spatial diffusion) is achieved around the profile scaling radius $r_s$. Factoring in diffusion, we find that the optimal angular region significantly increases. The optimal radius therefore depends both on the assumed dark matter profile and on the diffusion model, but in all cases we find that a radius of $\sim 6^{\prime}$ is a good choice.

We investigate whether we detect diffuse X-ray emission from these dwarfs above what we expect from the X-ray background.  First, we detect and exclude X-ray point sources; point sources are detected using the SAS wavelet detection routine, ewavelet, in a mosaiced 0.5-8 keV band image of all three detectors, and data within a $25^{\prime\prime}$ radius region of the source locations are excluded.  We then compare the detected flux within a $6^{\prime}$ radius aperture of the dwarf center to the expected X-ray background flux in that region of the detector by comparing to the EPIC blank-sky event files \citep{2007A&A...464.1155C}.  The blank-sky files are collected from all over the sky, while the dwarf galaxies considered here all lie outside of the galactic plane in regions of fairly low hydrogen column density.  We, therefore, use the tool BGSelector to create blank-sky files using only regions with similar galactic hydrogen column density to the dwarf galaxies; specifically, we filter on nH between $10^{20}$ and $5 \times 10^{20}$ cm$^{-2}$.  We then apply the same pattern, flag, and multi-band flare filtering to the background files as was used for the dwarf galaxy observations (Sec.~3).  Finally, we re-project the appropriate blank-field event files (medium filter for Fornax and thin filter for Ursa Minor and Carina) to match the sky position of the dwarf galaxy observations using the routine skycast.  Renormalizing the blank-field count rate in the source region (i.e. within $r=6^{\prime}$ excluding point source regions) using the ratio of the count rate in the blank-fields to that in the dwarf galaxy observations in an outer region of the detector (where the source flux is expected to be much lower), we find that none of the three dwarfs show significant diffuse X-ray emission above what is expected from the X-ray background.

As shown below, the non-detection of diffuse X-ray emission from dwarf galaxies places limits on the possible particle dark matter model.  We place an upper limit on the possible diffuse X-ray emission from each dwarf galaxy by assuming that all of the detected flux in the source region stems from the X-ray background and determining the necessary flux for a diffuse source to be detected at $3 \sigma$ above this background.  The derived flux limits are listed in Table 1 and range between $1 \times 10^{-5}$ and $2.5 \times 10^{-5}$ photons cm$^{-2}$ s$^{-1}$.
\begin{deluxetable}{lccccc}
\tablecaption{ Group Sample }
\tablewidth{0pt}
\tablecolumns{6}
\tablehead{
\colhead{Dwarf} & \colhead{R.A.} & \colhead{Decl.} & \colhead{Exposure (ks)} & \colhead{Flux Limit} & \colhead{Position Reference}\\
\colhead{} & \colhead{(J2000)} & \colhead{(J2000)} & \colhead{MOS1, MOS2, PN} & \colhead{(photons cm$^{-2}$ s$^{-1}$)}}
\startdata
Ursa Minor &15:09:11.3 &67:12:52 &11, 11, 7.8 &2.3E-5 &Cotton et al. 1999 \\
Fornax &02:39:52.0 &-34:30:49 &67, 69, 55 &1.0E-5 &Battaglia et al. 2006 \\
Carina &06:41:36.7 &-50:57:58 &19, 21, 13 &2.1E-5 &Lauberts 1982 \\
\enddata
\tablecomments{ Flux limits are listed for the 0.5-8 keV band for an aperture of $6^{\prime}$ radius. }
\end{deluxetable}

\subsection{Constraints on Dark Matter Models}\label{sec:constraints}

\begin{figure}
\mbox{\includegraphics[width=8cm]{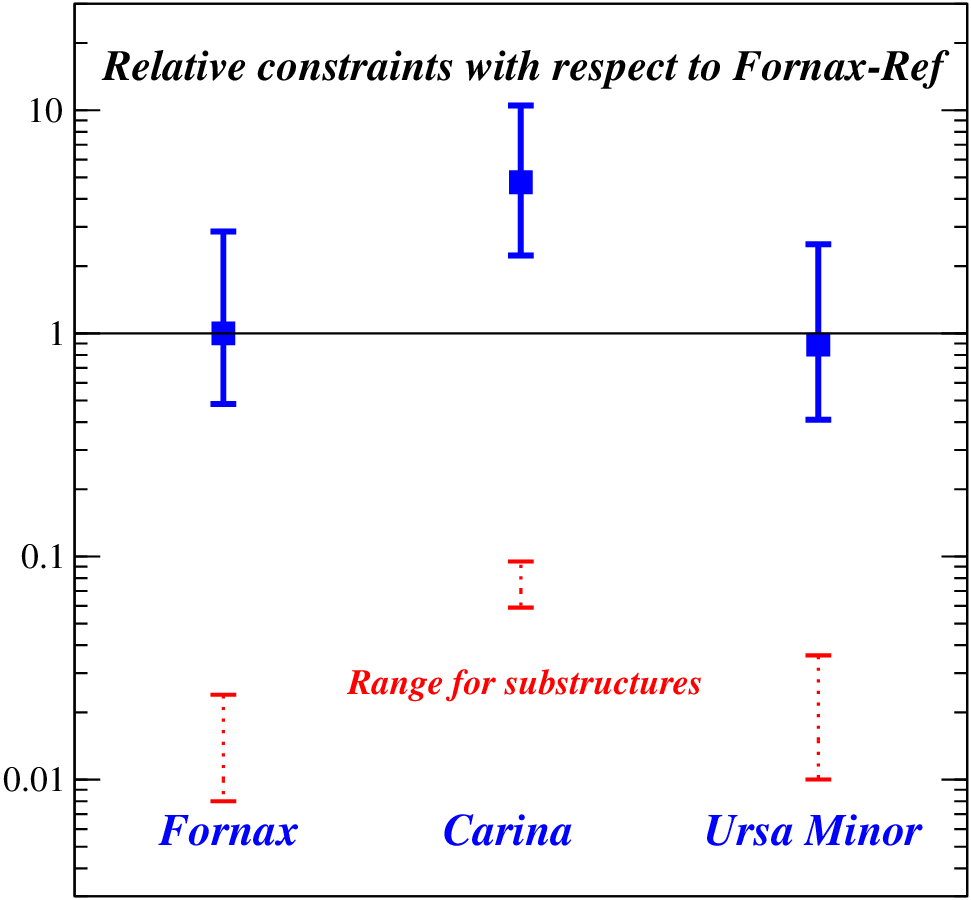}\quad\qquad\includegraphics[width=7.5cm]{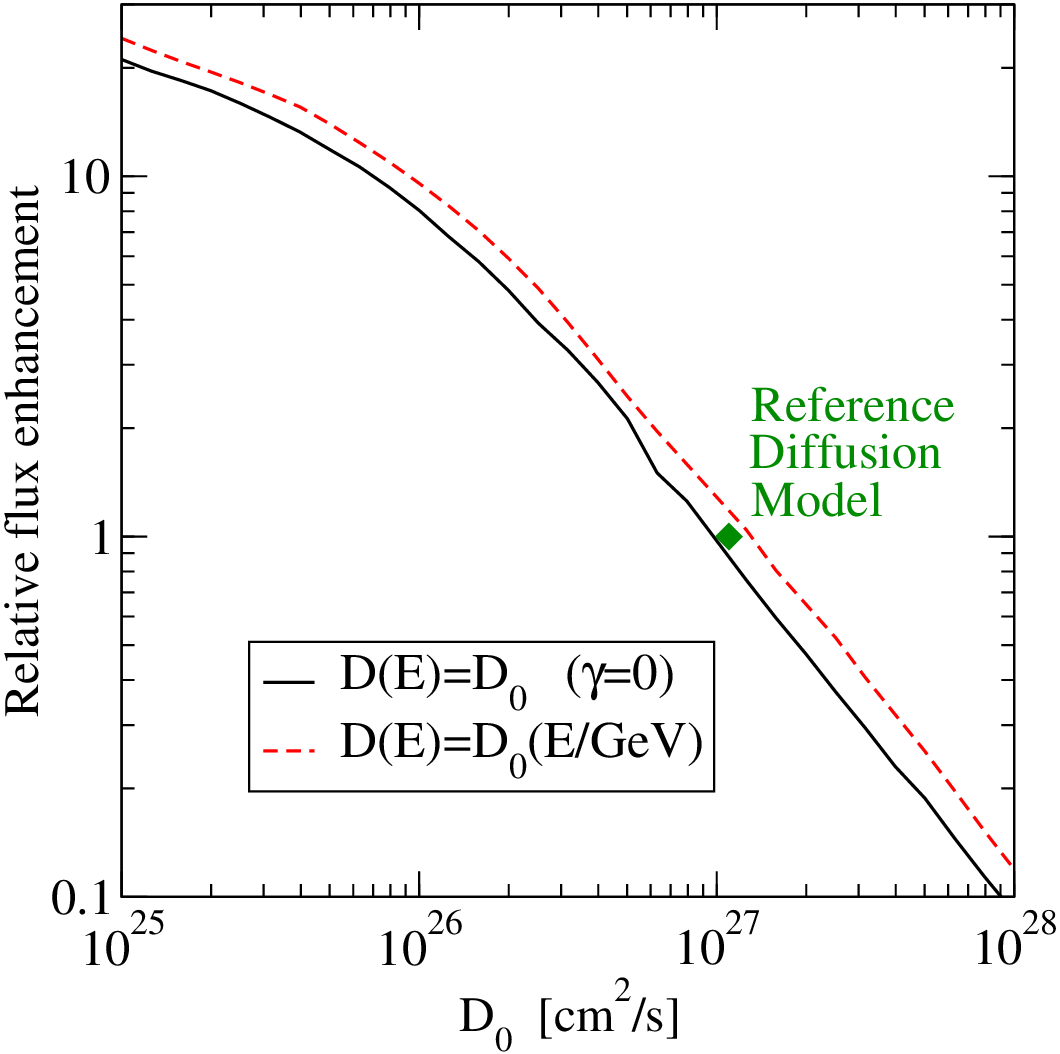}}
\caption{Left: a comparison of the constraints on WIMP models from the three dwarf galaxies under consideration here, normalized to the constraint we get for Fornax with the central reference value for the dark matter distribution. Stronger constraints correspond to smaller values on the $y$ axis. The blue solid lines indicate the ranges allowed, for an NFW profile, by dynamical data on the dark matter distribution of the various galaxies, assuming no substructures. The dotted red lines show how these ranges are affected accounting for dark matter sub-structures, according to the analysis of \citet{2007PhRvD..75h3526S}. Given the lack of consensus on the size of the boost factor from substructres, we regard the estimate quoted in \citet{2007PhRvD..75h3526S} as an indication of the maximal enhancement from substructures (see the text for more details on this point). Right: the effect on the integrated X-ray flux over the $[0.5,8]$ keV range of varying the diffusion coefficient $D_0$, normalized to our reference diffusion model (green diamond). The black curve assumes $\gamma=0$, while the red dashed line $\gamma=1$: at a given value of $D_0$, varying $\gamma$ between 0 and 1 shifts the X-ray flux from the lower to the upper line.\label{fig:compare}}
\end{figure}

We summarize and compare the diffuse X-ray flux limits we obtain for Fornax, Carina and Ursa Minor in the left panel of Fig.\ref{fig:compare}. To model the dark matter density distribution for these three dSph galaxies we employ NFW profiles, and follow the analysis of \cite{2007PhRvD..75h3526S} for the ranges of scale radii and densities allowed by dynamical data and CDM structure formation theoretical constraints.

Tab.~\ref{tab:profiles1} collects the reference, minimal and maximal values for the scaling density $\rho_s$, the scaling radius $r_s$, as well as the reference distance. Tab.~\ref{tab:profiles2} indicates instead the reference, minimal and maximal values for the angle-averaged line-of-sight integral $J$ for a solid angle  $\Delta\Omega\simeq10^{-5}$ sr, corresponding to an angle $\theta=6^\prime$. The quantity $J$ is defined as
\begin{equation}\label{eq:defj}
J=\frac{1}{\Delta\Omega}\int_0^{\Delta\Omega}\int_{\rm l.o.s.}\ \rho_{\rm DM}^2[r(s)]\ {\rm d}s.
\end{equation}
We express $J$ in units of $10^{23}\ {\rm GeV}^2\ {\rm cm}^{-5}$. Although we do consider the variation of the detector sensitivity with the offset angle (see Sec.~\ref{sec:data}), the values of $J$ give an idea of the normalization of the signal from different dwarfs \citep[see also][for further details]{2007PhRvD..75h3526S}. For instance, given the number $N_\gamma$ of photons produced per DM annihilation in a given energy band and for a certain particle dark matter model, the expected flux $\phi_\gamma$ in photons cm$^{-2}$ s$^{-1}$ in that energy band and from the solid angle $\Delta\Omega$ from a dSph with a normalization $J$ will simply be given by
\begin{equation}
\phi_\gamma=N_\gamma\frac{\langle\sigma v\rangle_0}{2\ m_{\rm DM}^2}\ J\ \frac{\Delta\Omega}{4\pi}.
\end{equation}
The last two columns of Tab~\ref{tab:profiles2} give the range for the substructure enhancement factor $B$ as obtained by \cite{2007PhRvD..75h3526S}. The rate of dark matter pair-annihilation in the presence of substructures is effectively boosted by the factor $B$. While the enhancement range is uncertain both because of the theoretical modeling and because the precise value of the cut-off scale for the smallest collapsed substructures depends critically on the particle dark matter model \citep[see e.g.][]{Profumosubs}, we refer here to the results reported by \cite{2007PhRvD..75h3526S} in their Fig.~5, and assume a small substructure cut-off, $M_{\rm cut}\sim 10^{-6}\ M_\odot$. The results of recent high-resolution N-body simulation of Milky-Way size galaxies reported in \cite{madau} indicate that the boost factor from substructures might be significantly smaller. For subhalo masses of the size of the dSph galaxies we consider here, and extrapolating the matter power spectrum with mass functions ${\rm d}N/{\rm d}M\propto M^{-\alpha}$, \cite{madau} finds boost factors ranging from $\sim1$ to $\sim 40$. The latter value refers to $M_{\rm cut}\sim 10^{-12}\ M_\odot$ and to $\alpha=2.1$, and should thus be regraded as an upper bound. The disruption of small substructures by stellar encounters can also play a relevant role, as discussed e.g. in \cite{stars}. It is beyond the scope of this analysis to assess the stability of the ranges quoted in \cite{2007PhRvD..75h3526S} against different N-body simulation results, extrapolations for the matter power spectrum at small scales and the mentioned particle dark matter uncertainties. We however warn the reader that the figures quoted in the last two columns of Tab~\ref{tab:profiles2} should be regraded as optimistic upper limits.

\begin{deluxetable}{lccccccc}
\tablecaption{ Dark Matter Profiles }
\tablewidth{0pt}
\tablecolumns{8}
\tablehead{
\colhead{Dwarf} & \colhead{$\rho_s^{\rm ref}$} & \colhead{$\rho_s^{\rm min}$} & \colhead{$\rho_s^{\rm max}$} & \colhead{$r_s^{\rm ref}$} & \colhead{$r_s^{\rm min}$} & \colhead{$r_s^{\rm max}$}& \colhead{$D$}}
\startdata
Ursa Minor & 7.5 & 7.35 & 7.85 & 0.2 & 0.067 & -0.033 & 66\\
Fornax & 7.6 & 7.35 & 7.9 & 0.05 & 0.067 & -0.067 & 138 \\
Carina & 7.8 & 7.5 & 8.0 & -0.3 & -0.23 & -0.36 & 101 \\
\enddata
\tablecomments{Range and reference values for the dark matter profiles of the three dSph galaxies under consideration. Columns 2, 3 and 4 indicate $\log_{10}[\rho_s/({M_\odot\ {\rm kpc}^{-3}})]$, columns 5, 6 and 7 quote $\log_{10}[r_s/{\rm kpc}]$ \citep{2007PhRvD..75h3526S}, and the last column is the reference dSph distance, in kpc.\label{tab:profiles1}}
\end{deluxetable}

\begin{deluxetable}{lccccc}
\tablecaption{ Dark Matter Density Squared Integrals and Substructure Enhancement Ranges}
\tablewidth{0pt}
\tablecolumns{6}
\tablehead{
\colhead{Dwarf} & \colhead{$J_{\rm ref}$} & \colhead{$J_{\rm min}$} & \colhead{$J_{\rm max}$} & \colhead{$B_{\rm low}$} & \colhead{$B_{\rm high}$} }
\startdata
Ursa Minor & 3.1 & 1.1 & 6.7 & 25 & 89 \\
Fornax & 1.2 & 0.42 & 2.49 & 50 & 159 \\
Carina & 0.53 & 0.24 & 1.13 & 50 & 80 \\
\enddata
\tablecomments{The line-of-sight integral of dark matter density squared, see Eq.~(\ref{eq:defj}), in units of $10^{23}\ {\rm GeV}^2\ {\rm cm}^{-5}$, and the range for the substructure boost factor $B$ as estimated in \citet{2007PhRvD..75h3526S}, for the three dSph galaxies under consideration here.\label{tab:profiles2}}
\end{deluxetable}

Fixing the dark matter density profile allows us to translate the X-ray flux limits given in the preceding section into actual constraints on the particle dark matter models. Our reference choices for the diffusion setup were specified above in Sec.~\ref{sec:multiw}, but the crucial dependence on the diffusion parameter $D$ will be further assessed here. Since it gives the strongest constraints on the X-ray flux, we choose to normalize our constraints to the Fornax dSph, with the reference dark matter setup specified in the second line of Tab.~\ref{tab:profiles1}. 

The left panel of Fig.~\ref{fig:compare} illustrates the relative strength of the constraints we get here for the three dSph under consideration, factoring in the uncertainty from the dark matter density distribution. A smaller number corresponds to a more stringent constraint, for instance on the pair-annihilation cross section for a given dark matter particle mass. The blue intervals correspond to the ranges on the normalization factor without substructures given in Tab.~\ref{tab:profiles2}. The dotted red ranges indicate the expected improvement on the constraints when the effect of substructures is included, according to the model of \cite{2007PhRvD..75h3526S}. The constraints obtained including substructures improve by the factors given in the last two columns of Tab.~\ref{tab:profiles2}. From the figure, we deduce that the impact on dark matter models of X-ray observations of Fornax is comparable to that obtained from data on Ursa Minor, because while the latter features a larger integrated dark matter density squared by a factor $\sim3$, its flare-free XMM exposure is much lower. Data from Carina provide constraints that are in the same ballpark of the other two dSph, but typically less stringent by factors of a few.

The right panel of Fig.~\ref{fig:compare} shows the effect on the flux of X-ray in the 0.5 to 8 keV band of varying the diffusion coefficient $D_0$ in the range between $10^{25}$ and $10^{28}\ {\rm cm}^2/$s. We normalize the flux to that obtained in our reference diffusion setup, and show the lines corresponding to $\gamma=0$ (solid black) and $\gamma=1$ (red dashed). Intermediate values of $\gamma$ fall between the two lines. Recall that the reference diffusion setup features $\gamma=0.7$. The proximity of the two lines indicates that the precise value of $\gamma$ is much less critical to the X-ray emission from $e^+e^-$ produced by dark matter annihilation than the value of $D_0$. As expected, smaller values of the diffusion coefficient induce a smaller loss of energetic $e^+e^-$, and, eventually, for smaller and smaller $D_0$ the curves will converge to the value corresponding to a scenario where diffusion can be totally neglected. In the range we explored, diffusion can lead to a suppression of the signal by a factor $\sim10$ for larger values of $D_0$, or to enhancements by more than a factor 20 for smaller values of $D_0$ compared to our reference setup.

\begin{figure}
\begin{center}
\mbox{\includegraphics[width=10cm]{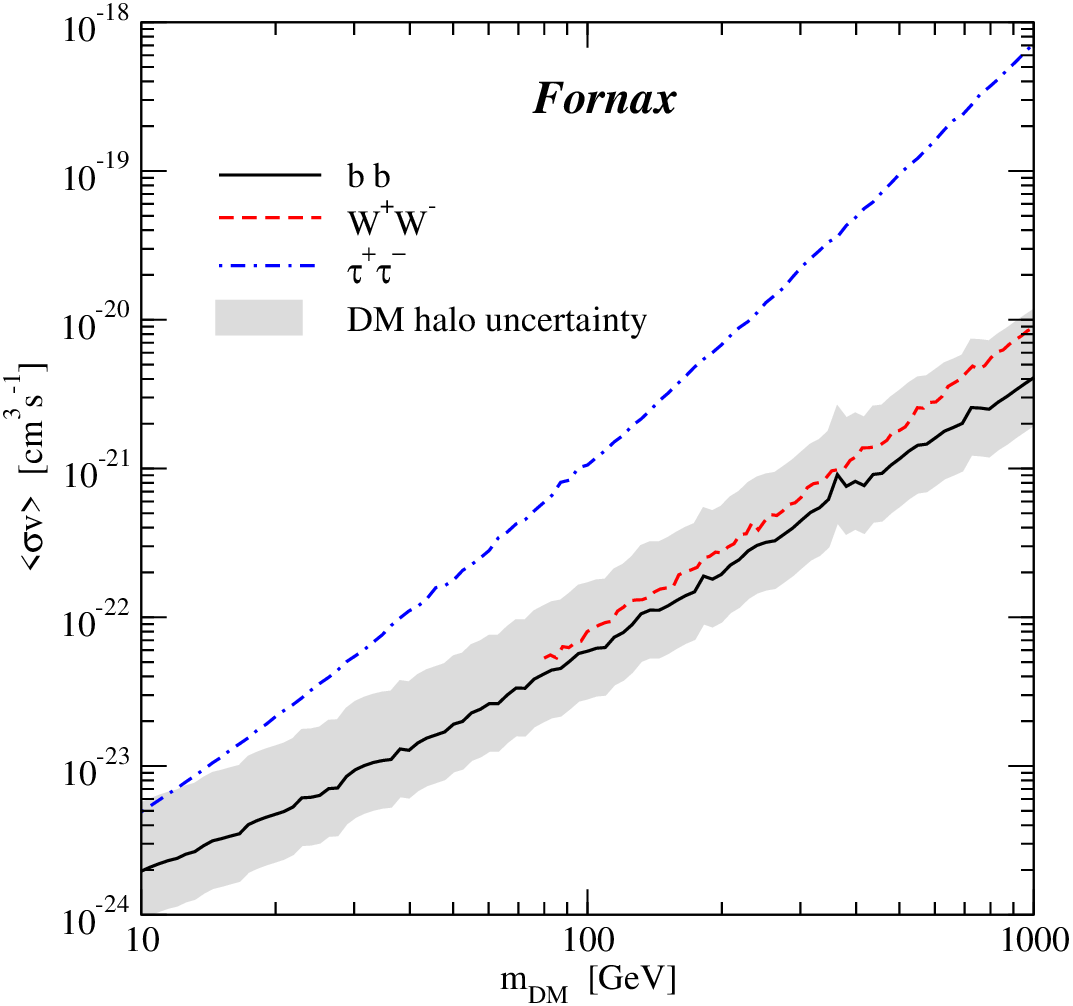}}
\end{center}
\caption{A comparison of the constraints on the WIMP mass versus pair-annihilation cross section plane for different final states: $b\bar b$, $W^+W^-$ and $\tau^+\tau^-$, neglecting dark matter substructures, and for the case of Fornax. The gray band indicates the uncertainty (for definiteness around the $b\bar b$ line) in the dark matter profile modeling.\label{fig:fs}}
\end{figure}

As alluded to above, given the dark matter density distribution, and fixing the dark matter mass and the $e^+e^-$ injection spectrum (via the dominant pair-annihilation final state), we can determine the maximal values of the pair-annihilation cross section $\langle\sigma v\rangle_0$ allowed by X-ray data. We carry out this exercise in Fig.~\ref{fig:fs} assuming {\em no substructure enhancement}, and for the reference dark matter and diffusion setups, for the benchmark case of Fornax. The solid, dashed and dot-dashed lines correspond to the three dominant pair-annihilation final states, respectively $b\bar b$, $W^+W^-$ and $\tau^+\tau^-$. We also show the uncertainty band (with respect to the $b\bar b$ final state) stemming from the determination of the dark matter density distribution, neglecting the effect of substructures, and the possibility of assuming density distribution profiles different from a NFW profile. Under these very conservative assumptions we are able to constrain interesting values of the pair annihilation cross section, particularly in the light WIMP mass range \citep[see e.g.][for a discussion of the phenomenology of light ($m_{\rm DM}\ll 1$ GeV) neutralinos]{profumolight}. Models producing a softer $e^+e^-$ spectrum, such as $b\bar b$ are more strongly constrained than models featuring a harder spectrum (such as $\tau^+\tau^-$, where energetic electrons and positrons are produced in both the leptonic and in the hadronic $\tau$ decays). Interestingly, the final states that are usually dominant in supersymmetric models, $b\bar b$ and $W^+W^-$, give comparable constraints, well within the dark matter profile uncertainties. Since pair-annihilation modes such as $\tau^+\tau^-$ occur in supersymmetric models with a significant branching ratio only if, for instance, the scalar partners of the $\tau$ are light, and are in any case subdominant with respect to the decays into heavy quarks, it is legitimate here to use, for the case of supersymmetry, what we obtain for a $b\bar b$ final state.

\begin{figure}
\mbox{\includegraphics[width=15cm]{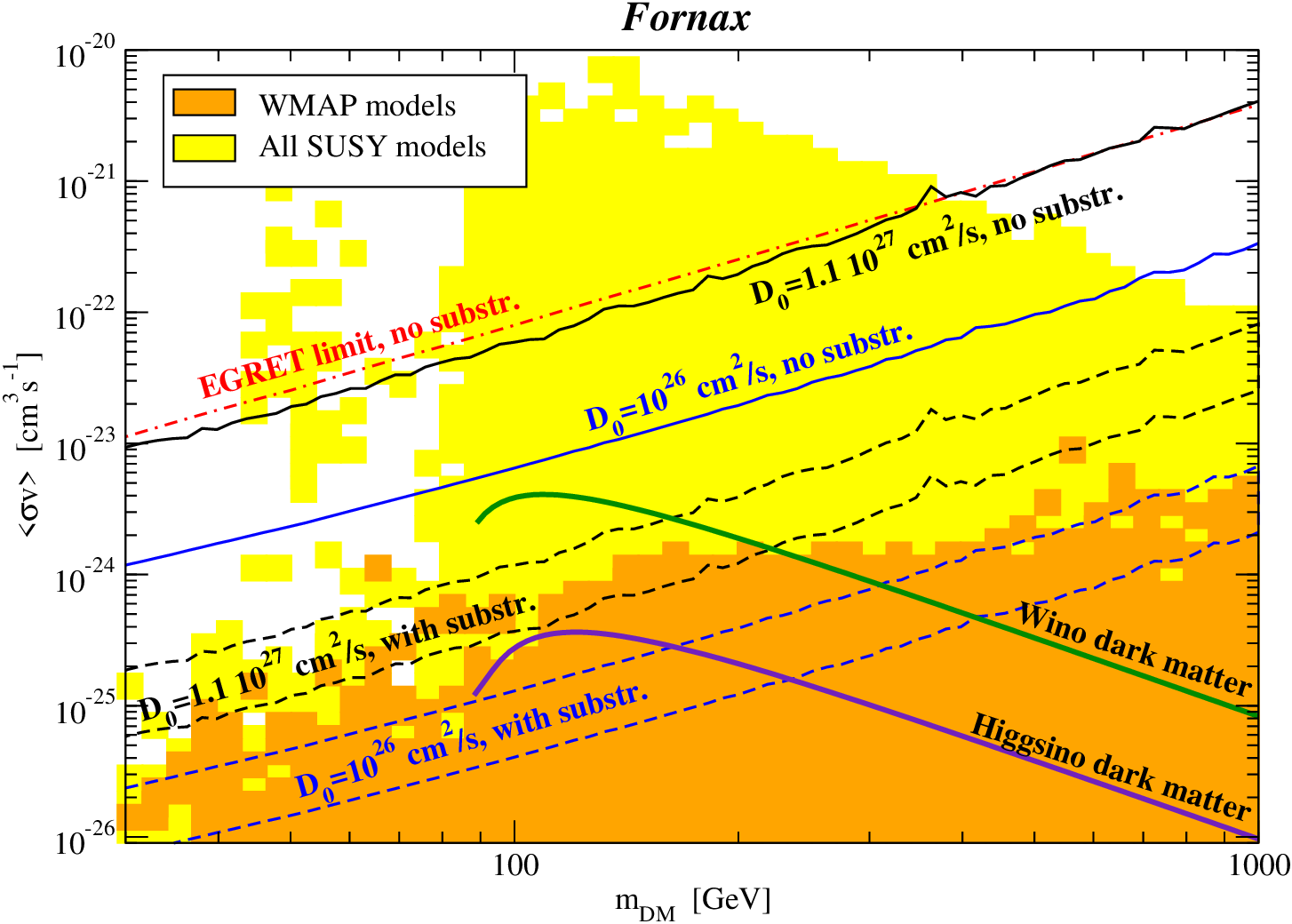}}
\caption{Overview of the constraints from X-ray observations of Fornax on supersymmetric WIMP dark matter models. We show the limits obtained using our conservative reference diffusion setup and a diffusion coefficient $D_0=10^{26}\ {\rm cm}^2/{\rm s}$. The solid lines correspond to the case of no substructures, while the dashed lines indicate the range where one could expect the limit to be set when substructures are included. The yellow area corresponds to values of the dark matter mass and annihilation cross section found for neutralino models within the minimal supersymmetric extension of the Standard Model. The orange area indicates those supersymmetric models that also produce a thermal relic neutralino abundance in agreement with the inferred cold dark matter density. The solid green and indigo lines locate the predictions for ``vanilla'' Wino and Higgsino dark matter models.\label{fig:fornax}}
\end{figure}

We give an overview of the constraints we can place on supersymmetric models in Fig.~\ref{fig:fornax}, for different assumptions on the diffusion and the dark matter density distribution for the Fornax dSph galaxy. We again illustrate our constraints in the $(m_{\rm DM}, \langle\sigma v\rangle_0)$ parameter space, and assume that the spectrum of $e^+e^-$ is close enough to a $b\bar b$ final state. Dark shaded (orange) regions correspond to values for $(m_{\rm DM}, \langle\sigma v\rangle_0)$ populated by supersymmetric models with a thermal relic abundance consistent with the inferred cold dark matter abundance \citep[see e.g.][]{wmap}. Light shaded (yellow) points indicate those portions of the parameter space where in a standard $\Lambda$CDM cosmology the thermal relic abundance of dark matter is under-produced. In this case, we assume that either non-thermal dark matter production occurs, or that the Universe underwent a non-standard expansion rate phase prior to Big Bang Nucleosynthesis and in particular around the WIMP freeze-out \citep[see e.g.][for the discussion of one such non-standard cosmological setup involving a Quintessence field describing dark energy]{quint}. In short: the dark shaded region indicates where the WIMP pair-annihilation cross section is compatible with a ``vanilla'' thermal freeze-out scenario, while the light shaded region gives models with a more extreme annihilation rate, compatible with the cold dark matter density provided additional assumptions are made on the WIMP production mechanism beyond standard thermal generation. We refer the reader to \cite{2005PhRvD..72j3521P} for details on the scan of the parameter space of the minimal supersymmetric extension of the Standard Model employed to obtain the shaded regions shown in Fig.~\ref{fig:fornax}.

The upper solid black line reproduces the constraint shown in Fig.~\ref{fig:fs}, and assumes no substructures and a conservative value for the diffusion coefficient (i.e.~the reference model matching the median diffusion coefficient for the Milky Way). We compare the X-ray constraints with the EGRET limit on the gamma-ray flux, indicated with the nearby dot-dashed red line. For EGRET, we assume a point-source sensitivity of around $10^{-8} {\rm cm}^{-2}{\rm s}^{-1}$, and an angular acceptance of 1 deg \citep{egretpaper}. We see that X-ray limits are at least as constraining as existing gamma-ray limits, even with conservative assumptions on the $e^+e^-$ diffusion setup. The blue solid line roughly one order of magnitude below the two lines discussed above indicates the constraints for a reduced diffusion coefficient, $D_0=10^{26}\ {\rm cm}^2/$s, our estimate for the diffusion coefficient scaled down from the Milky Way to the scale of a dSph galaxy (Sec.~2). With this diffusion setup, existing X-ray limits are one order of magnitude better than existing gamma-ray limits, and can be in some cases competitive with the GLAST anticipated performance (a similar result was recently obtained for the case of the galactic center by \cite{ullioregis}, although the dominant X-ray production mechanism in that case is synchrotron radiation and not Inverse Compton scattering).

Adding substructures to the description of the dark matter density pushes the limits well within the area where WIMPs are produced thermally in the early universe, even with a conservative diffusion setup. This statement is substantiated in Fig.~\ref{fig:fornax} by the two sets of black and blue dashed lines, corresponding, respectively for $D_0=1.1\times 10^{27}\ {\rm cm}^2/$s and $D_0=10^{26}\ {\rm cm}^2/$s, to the substructure boost factors indicated in Tab.~\ref{tab:profiles2}. For reference, we indicate with solid green and indigo lines the expectation for dominantly wino-like and higgsino-like lightest neutralinos. The case of wino-like dark matter is ubiquitous in so-called anomaly-mediated supersymmetry-breaking scenarios, where the SU(2) soft-supersymmetry breaking gaugino mass is much smaller than its U(1) hyper-charge counterpart \citep[see e.g.][]{amsb}. Wino dark matter is constrained, in the most extreme setup of substructure enhanced and suppressed diffusion, up to masses around 400 GeV. Higgsino dark matter is also ubiquitous in several models of supersymmetry breaking \citep[see e.g.][]{marcsreview}, including the focus point region of minimal supergravity \citep[see e.g.][]{focusp}.

An important point we wish to make here is that the constraints we show here are consistent with all available particle physics bounds on dark matter. In particular, it was shown in \cite{ophiuchus} (see the upper panel of Fig.~2) that conservative limits in the $(m_{\rm DM}, \langle\sigma v\rangle_0)$ plane for $10\lesssim m_{\rm DM}/{\rm GeV}\lesssim 1000$ range between $\langle\sigma v\rangle_0/({\rm cm}^3{\rm s}^{-1})\lesssim 10^{-19}$ and $\langle\sigma v\rangle_0/({\rm cm}^3{\rm s}^{-1})\lesssim 10^{-21}$. Given uncertainties on cosmic ray propagation in the galaxy, these limits were obtained assuming a cored dark matter density profile and considering gamma-ray data from EGRET and H.E.S.S. on the galactic center region, according to the analysis of \cite{2004APh....21..267C} and of \cite{2005PhRvD..72j3521P}. Limits on the dark matter pair annihilation cross section were also obtained by \cite{2007PhRvL..99w1301B} by comparing the cosmic diffuse neutrino signal that would result from dark matter pair annihilation and comparing it to the measured terrestrial atmospheric neutrino background. The limits quoted by \cite{2007PhRvL..99w1301B} are also at the level of $\langle\sigma v\rangle_0/({\rm cm}^3{\rm s}^{-1})\lesssim 2\times 10^{-21}$ in the $10\lesssim m_{\rm DM}/{\rm GeV}\lesssim 1000$ range. In short, the limits we quote here fall in a region of the particle dark matter parameter space which is not ruled out by existing data on indirect detection.

To summarize, limits on the diffuse X-ray flux from dSph galaxies effectively constrain WIMP dark matter models. The strength of these constraints depends sensitively on the dark matter density distribution (and more specifically on the effect of dark matter substructures) and on the diffusion model. Less conservative assumptions on either one of those points pushes the constraints into interesting regions of the parameter space of particle dark matter models (see Fig.~\ref{fig:fornax}). We showed that currently available X-ray and gamma-ray data from nearby dSph galaxies put comparable constraints on particle dark matter indirect detection even for very conservative diffusion setups; a suppression of cosmic ray diffusion appropriate for a dSph scale galaxy makes X-ray data more sensitive to particle dark matter pair annihilation by about one order of magnitude compared to gamma rays. In the next section we delineate a comparison between the soon to be launched gamma-ray telescope GLAST and future X-ray missions in terms of their sensitivity on extra-galactic particle dark matter searches.

\section{Summary and Discussion}\label{sec:discussion}

We pointed out that local dSph galaxies are an ideal environment for particle dark matter searches with X-rays. We described how X-rays are produced as secondary radiation in Inverse Compton scattering off cosmic microwave background photons of electrons and positrons resulting from particle dark matter annihilation. The resulting spectrum is only mildly dependent on the details of the particle dark matter model (the dark matter mass and the dominant final state into which it pair annihilates), and it is, generically, hard (spectral index smaller than $\sim 1.5$). The normalization of the emission depends on (1) the particle dark matter pair annihilation rate, (2) the diffusion setup, and (3) the dark matter density distribution. For reasonable choices of these three {\em a priori} unknown inputs of the problem, the X-ray emission is potentially within reach of current X-ray detectors. Interestingly enough, the shape of the spectral energy distribution indicates that for dSph galaxies X-rays have a comparable, if not better, sensitivity to indirect dark matter detection than gamma rays.

We used XMM-Newton archival data on three Local Group dSph galaxies, Ursa Minor, Fornax and Carina, to search for the diffuse X-ray emission expected from dark matter annihilation. We studied the optimal energy and radial range to search for this type of emission, and concluded that for XMM-Newton and for the dSph galaxies under investigation these correspond to an energy band between 0.5 and 8 keV and to a radius of around $6^\prime$. We do not find any significant signal over background, and this, in turn, was turned into constraints on particle dark matter models. The best constraints result from both the Fornax and the Ursa Minor observations, while data from Carina result in bounds that are a factor of a few weaker.  Ursa Minor features the largest dark matter density, making it the best candidate target, but has the shortest usable XMM exposure.

In determining the impact on particle dark matter searches of our X-ray constraints, we pointed out the uncertainties resulting from the modeling of cosmic ray diffusion processes, and from the dark matter distribution. In particular, including dark matter substructures can boost our constraints significantly. We phrase the bounds we obtain in terms of the dominant dark matter annihilation final states. For those final states relevant for specific dark matter models, such as supersymmetry, the constraints on the mass versus pair annihilation plane are very similar. We then proceeded to examine how X-ray constraints on particle dark matter annihilation in local dSph galaxies limit the available parameter space of supersymmetric dark matter. In the most conservative setup only models with rather large annihilation cross sections are excluded. Assuming a smaller diffusion coefficient, or factoring in the effect of dark matter substructures, our constraints fall well within the interesting region where the supersymmetric dark matter can be a thermal relic from the early universe. Also, we were able to set limits on particular supersymmetric dark matter scenarios, such as Wino or Higgsino lightest neutralino dark matter.

An important result of the present analysis is that even assuming a conservative diffusion setup the sensitivity of X-rays and of gamma rays to particle dark matter annihilation in dSph galaxies are comparable. This fact has two-fold implications: on the one hand, if longer observations of dSph galaxies were carried out with existing telescopes, it is possible that the first astronomical signature of particle dark matter annihilation would come from X-rays. Secondly, should a signature be detected in the future with gamma-ray telescopes, it would be extremely important to confirm the nature of the signal via X-ray observations.

In this respect, it is relevant to comment here on how future gamma-ray and X-ray telescopes will improve indirect dark matter searches through observations of nearby dSph galaxies. The LAT instrument on-board the soon to be launched GLAST satellite will extend the gamma-ray energy range available to EGRET, with tremendously increased effective area and energy as well as angular resolution. GLAST will be an ideal telescope to search for dark matter annihilation in dSph galaxies. Assuming a diffuse background flux of $1.5\times 10^{-5}$ photons cm${}^{-2}$ s$^{-1}$ sr$^{-1}$ integrated above 0.1 GeV, and an effective spectral index in the gamma-ray band of 2.1, we find that the GLAST LAT sensitivity\footnote{http://www-glast.slac.stanford.edu/software/IS/glast\_lat\_performance.htm} from 5 years of data will improve over the EGRET point-source sensitivity by large factors.  In the mass versus pair annihilation cross section plane, and assuming a soft gamma ray spectrum ($b\bar b$), GLAST will improve over EGRET by factors ranging between $\sim10$ and $\sim100$, the first corresponding to a light $m_{\rm DM}\sim10$ GeV dark matter particle, and the latter to a heavy one ($m_{\rm DM}\sim1000$ GeV). Assuming a harder gamma-ray spectrum, as appropriate for other dark matter models \citep[e.g. universal extra dimensions,][]{ued}, the GLAST performance will be factors between 30 and 300 better than EGRET. A signal of dark matter pair annihilations in gamma-rays appears therefore very promising with GLAST. If detected, such a source would need to be confirmed in its nature, and using a multi-wavelength approach is one of the most promising strategies. 

Future X-ray telescopes, like Constellation-X and XEUS, will also have greatly increased effective areas with respect to current instruments. Using the currently available projections for the effective areas and backgrounds of these telescopes\footnote{http://sci.esa.int/science-e/www/object/index.cfm?fobjectid=42273 \\and http://constellation.gsfc.nasa.gov/resources/response\_matrices/index.html}, we estimate that the X-ray limits (0.5-8 keV band) placed by a 100 ksec observation of Ursa Minor with Constellation-X or XEUS would improve over the limits placed in this paper by factors of roughly 35 and 70, respectively.  Thus even for a conservative diffusion model, the future generation of X-ray telescopes will place similar constraints on dark matter annihilation from dwarfs as GLAST, stronger constraints at particle masses below a few hundred GeV. In addition, a signal from GLAST could be confirmed with X-ray observations. 

Observations at other wavelengths will also be of great relevance to identify particle dark matter and its properties. In particular, a diffuse radio signal should also be part of the multiwavelength yield of particle dark matter annihilation. However, the level of the radio emission is crucially dependent upon the average magnetic field, which adds further uncertainties both in setting constraints and in understanding the nature of particle dark matter, should a signal be detected. Observations in the hard X-ray band would also be useful; however, as opposed to cluster of galaxies, where the effect of diffusion on the dark matter multi-wavelength SED is typically mild \citep[see][]{colacoma,ophiuchus}, in dSph galaxies high energy electrons and positrons escape more efficiently from the diffusive region, suppressing the hard X-ray emission. 

\cite{2007arXiv0709.1510S} recently estimated the gamma-ray flux from dark matter annihilation in newly discovered, extremely low luminosity and dark matter dominated Milky Way dwarf galaxies. The expected gamma-ray flux from these nearby galaxies can be larger than that from previously known dwarfs, depending on the so far poorly known dark matter density distribution. Our results indicate that these objects would also potentially be excellent targets for particle dark matter searches with X-ray observations.

In short, we showed that X-rays can play an important role in exploring the nature of particle dark matter and in pinpointing its properties. This role is complementary, but not subsidiary, to searches with gamma rays, and we believe very exciting results at both frequencies might be just around the corner.

\acknowledgments
We acknowledge useful discussions with Fiorenza Donato and Piero Ullio on cosmic ray diffusion. T.E.J. is grateful for support from the Alexander F. Morrison Fellowship, administered through the University of California Observatories and the Regents of the University of California.


\clearpage

\begin{thebibliography}{99}

\bibitem[Aloisio et al.(2004)]{aloisio0402588} \mbox{Aloisio, R., Blasi, P., 
\& Olinto, A.~V.\ 2004, Journal of Cosmology and Astro-Particle Physics, 5, 7} 

\bibitem[Baer \& Profumo(2005)]{dbar} Baer, H., \& Profumo, S.\ 2005, Journal of Cosmology and Astro-Particle Physics, 12, 8 

\bibitem[Baer et al.(2005)]{focusp} Baer, H., Krupovnickas, 
T., Profumo, S., \& Ullio, P.\ 2005, Journal of High Energy Physics, 10, 20 

\bibitem[Baltz \& Edsj{\"o}(1999)]{1999PhRvD..59b3511B} Baltz, E.~A., \& Edsj{\"o}, J.\ 1999, \prd, 59, 023511 

\bibitem[Baltz \& Wai(2004)]{baltzwai} Baltz, E.~A., \& Wai, L.\ 2004, \prd, 70, 023512 

\bibitem[Battaglia et al.(2006)]{battaglia} Battaglia, G., et al.\ 2006, \aap, 459, 423

\bibitem[Beacom et al.(2007)]{2007PhRvL..99w1301B} Beacom, J.~F., Bell, 
N.~F., \& Mack, G.~D.\ 2007, Physical Review Letters, 99, 231301

\bibitem[Bertone et al.(2001)]{bertone0101134} Bertone, G., Sigl, G., \& Silk, J.\ 2001, \mnras, 326, 799 

\bibitem[Bertone et al.(2005)]{hooperreview} Bertone, G., Hooper, 
D., \& Silk, J.\ 2005, \physrep, 405, 279 

\bibitem[Birkedal et al.(2006)]{lhm} Birkedal, A., Noble, A., Perelstein, M., \& Spray, A.\ 2006, \prd, 74, 035002 

\bibitem[Bringmann \& Salati(2007)]{2007PhRvD..75h3006B} Bringmann, T., \& Salati, P.\ 2007, \prd, 75, 083006 

\bibitem[Brun et al.(2007)]{2007PhRvD..76h3506B} Brun, P., Bertone, G., Lavalle, J., Salati, P., \& Taillet, R.\ 2007, \prd, 76, 083506 

\bibitem[Carter \& Read(2007)]{2007A&A...464.1155C} Carter, J.~A., \& Read, A.~M.\ 2007, \aap, 464, 1155

\bibitem[Cesarini et al.(2004)]{2004APh....21..267C} Cesarini, A., Fucito, 
F., Lionetto, A., Morselli, A., 
\& Ullio, P.\ 2004, Astroparticle Physics, 21, 267 

\bibitem[Colafrancesco et al.(2006)]{colacoma} Colafrancesco, S., Profumo, S., \& Ullio, P.\ 2006, \aap, 455, 21 

\bibitem[Colafrancesco et al.(2007a)]{coladraco} Colafrancesco, 
S., Profumo, S., \& Ullio, P.\ 2007a, \prd, 75, 023513 

\bibitem[Colafrancesco et al.(2007b)]{colabullet} Colafrancesco, S., de Bernardis, P., Masi, S., Polenta, G., \& Ullio, P.\ 2007b, \aap, 467, L1 

\bibitem[Cotton et al.(1999)]{cotton} Cotton, W.~D., Condon, J.~J., \& Arbizzani, E.\ 1999, \apjs, 125, 409

\bibitem[Donato et al.(2004)]{2004PhRvD..69f3501D} Donato, F., Fornengo, 
N., Maurin, D., Salati, P., \& Taillet, R.\ 2004, \prd, 69, 063501 

\bibitem[Finkbeiner(2004)]{haze1} Finkbeiner, D.~P.\ 2004, 
ArXiv Astrophysics e-prints, arXiv:astro-ph/0409027 

\bibitem[Gondolo(2000)]{gondolo} Gondolo, P.\ 2000, Physics Letters B, 494, 181 
\bibitem[Hartman et al.(1999)]{egretpaper} Hartman, R.~C., et al.\ 1999, \apjs, 123, 79 

\bibitem[Hooper \& Profumo(2007)]{ued} Hooper, D., \& Profumo, S.\ 2007, \physrep, 453, 29 

\bibitem[Hooper et al.(2007)]{haze2} Hooper, D., Finkbeiner, 
D.~P., \& Dobler, G.\ 2007, \prd, 76, 083012 

\bibitem[Hooper et al.(2008)]{haze3} Hooper, D., Zaharijas, 
G., Finkbeiner, D.~P., \& Dobler, G.\ 2008, \prd, 77, 043511 

\bibitem[Hooper(2008)]{2008arXiv0801.4378H} Hooper, D.\ 2008, ArXiv 
e-prints, 801, arXiv:0801.4378 

\bibitem[Jungman et al.(1996)]{marcsreview} Jungman, G., 
Kamionkowski, M., \& Griest, K.\ 1996, \physrep, 267, 195 

\bibitem[Klein et al.(1992)]{klein} Klein, U., Giovanardi, C., Altschuler, D.~R., \& Wunderlich, E.\ 1992, \aap, 255, 49 

\bibitem[Komatsu et al.(2008)]{wmap} Komatsu, E., et al.\ 
2008, ArXiv e-prints, 803, arXiv:0803.0547 

\bibitem[Kuhlen et al.(2008)]{madau} Kuhlen, M., Diemand, J., \& Madau, P.\ 2008, ArXiv e-prints, 805, arXiv:0805.4416 

\bibitem[Lauberts(1982)]{lauberts} Lauberts, A.\ 1982, Garching: European Southern Observatory (ESO), 1982

\bibitem[Lazarian \& Cho(2004)]{2004Ap&SS.289..307L} Lazarian, A., \& Cho, J.\ 2004, \apss, 289, 307 

\bibitem[Mateo(1998)]{1998ARA&A..36..435M} Mateo, M.~L.\ 1998, \araa, 36, 435 

\bibitem[Moroi \& Randall(2000)]{amsb}  Moroi, T., \& Randall, L.\ 2000, Nuclear Physics B, 570, 455 

\bibitem[Narayan \& Medvedev(2001)]{2001ApJ...562L.129N} Narayan, R., \& Medvedev, M.~V.\ 2001, \apjl, 562, L129 

\bibitem[Navarro et al.(1997)]{nfwpaper} Navarro, J.~F., Frenk, 
C.~S., \& White, S.~D.~M.\ 1997, \apj, 490, 493 

\bibitem[Nevalainen et al.(2005)]{neva} Nevalainen, J., Markevitch, M., \& Lumb, D. 2005, ApJ, 629, 172

\bibitem[Profumo \& Ullio(2003)]{quint} Profumo, S., \& Ullio, P.\ 2003, Journal of Cosmology and Astro-Particle Physics, 11, 6 

\bibitem[Profumo \& Ullio(2004)]{amprofumo} Profumo, S., \& Ullio, P.\ 2004, Journal of Cosmology and Astro-Particle Physics, 7, 6 

\bibitem[Profumo(2005)]{2005PhRvD..72j3521P} Profumo, S.\ 2005, \prd, 72, 
103521 

\bibitem[Profumo et al.(2006)]{Profumosubs} Profumo, S., Sigurdson, 
K., \& Kamionkowski, M.\ 2006, Physical Review Letters, 97, 031301 

\bibitem[Profumo(2008a)]{ophiuchus} Profumo, S.\ 2008a, ArXiv 
e-prints, 801, arXiv:0801.0740 [to appear in \prd]

\bibitem[Profumo(2008b)]{profumolight} Profumo, S, 2008b, submitted to \prd

\bibitem[Regis \& Ullio(2008)]{ullioregis} Regis, M., \& Ullio, P.\ 2008, ArXiv e-prints, 802, arXiv:0802.0234 

\bibitem[Strigari et al.(2007a)]{2007PhRvD..75h3526S} Strigari, L.~E., 
Koushiappas, S.~M., Bullock, J.~S., 
\& Kaplinghat, M.\ 2007a, \prd, 75, 083526 

\bibitem[Strigari et al.(2007b)]{2007arXiv0709.1510S} Strigari, L.~E., 
Koushiappas, S.~M., Bullock, J.~S., Kaplinghat, M., Simon, J.~D., Geha, M., 
\& Willman, B.\ 2007b, ArXiv e-prints, 709, arXiv:0709.1510 

\bibitem[Zakamska \& Narayan(2003)]{2003ApJ...582..162Z} Zakamska, N.~L., \& Narayan, R.\ 2003, \apj, 582, 162 

\bibitem[Zhao et al.(2007)]{stars} Zhao, H., Hooper, D., 
Angus, G.~W., Taylor, J.~E., \& Silk, J.\ 2007, \apj, 654, 697 


\end{thebibliography}
\end{document}